\documentclass[11pt, leqno]{article}

\usepackage{amsmath, amsthm, amssymb, setspace, fullpage, graphicx, cite,color}
\long\def\old#1{}
\def\E{\mathbb{E}}
\def\B{{\mathcal{B}}}
\def\P{\mathbb{P}}

\title{Delay Stability Regions of the Max-Weight Policy under Heavy-Tailed Traffic \footnote{This work was supported by NSF Grants CNS-0915988 and CCF-0728554, and ARO MURI Grant W911NF-08-1-0238. The material in this paper was presented in part at the $31^{st}$ IEEE INFOCOM Conference, Orlando, FL, USA.}%
}

\author{Mihalis G. Markakis, Eytan Modiano, and John N. Tsitsiklis
\thanks{The authors are with the Laboratory for Information and Decision Systems, at the Massachusetts Institute of Technology, Cambridge, MA, USA. Emails: \{mihalis,modiano,jnt\}@mit.edu.}%
}

\begin{document}

\date{}

\maketitle

\begin{abstract}

We carry out a delay stability analysis (i.e., determine conditions under which expected steady-state delays at a queue are finite) for a simple 3-queue system operated under the Max-Weight scheduling policy, for the case where one of the queues is fed by heavy-tailed traffic (i.e, when the number of arrivals at each time slot has infinite second moment). This particular system exemplifies an intricate phenomenon whereby heavy-tailed traffic at one queue may or may not result in the delay instability of another queue,  depending on the arrival rates.

While the ordinary stability region (in the sense of convergence to a steady-state distribution) is straightforward to determine, the determination of the delay stability region is more involved: 
(i) we use ``fluid-type'' sample path arguments, combined with renewal theory, to prove delay instability outside a certain region; (ii) we use a piecewise linear Lyapunov function to prove delay stability in the interior of that same region; (iii) as an intermediate step in establishing delay stability, we show that the expected workload of a stable M/GI/1 queue scales with time as $\mathcal{O}(t^{1/(1+\gamma)})$, assuming that service times have a finite $1+\gamma$ moment, where $\gamma \in (0,1)$.

\end{abstract}


\bigskip\section{Introduction}

\par 
We consider a simple 3-queue system operated under the Max-Weight scheduling policy, for the case where one of the queues is fed by heavy-tailed traffic (i.e, when the number of arrivals at each time slot has infinite second moment). The performance metric that we are interested in is {\bf delay stability}\/: a queue is delay stable if the expected steady-state delay in that queue is finite, and delay unstable otherwise.

 Under the Max-Weight scheduling policy, the notions of stability and delay stability coincide when all primitive stochastic processes are light-tailed; that is, all queues are delay stable for all arrival rate vectors within the stability region of the system (see, e.g., \cite{TE92}). On the other hand, for a system of parallel queues served by a single server, it is known that if one of the arrival streams is heavy-tailed, then all queues are delay unstable, for  all positive arrival rates \cite{MMT09}. The system that we study in this paper exhibits an intermediate behavior: we show that in the presence of heavy-tailed traffic, there may be nontrivial parts of the stability region where a queue is delay stable, and nontrivial parts of the stability region where the same queue is delay unstable. This exemplifies an intricate ``delay instability propagation'' whereby heavy-tailed traffic at one queue may or may not result in the delay instability of another queue, depending on the numerical values of the arrival rates.

\par The study of queueing systems with heavy-tailed traffic is motivated by significant evidence that traffic in real-world networks exhibits strong correlations and statistical self-similarity over different time scales. This observation was first made by Leland \emph{et al.} \cite{LTWW94} through an analysis of Ethernet traffic traces. Subsequent empirical studies have documented this phenomenon in other networks, while accompanying theoretical studies have associated it with heavy-tailed arrival processes; see \cite{PW00} for an overview. At the same time, the performance analysis of Max-Weight scheduling is of independent interest, with a long history and rich literature, see, e.g., \cite{AKRSVW04, DL05, S04, TE92}. The feature that makes the Max-Weight policy appealing is its throughput optimality, i.e., its ability to stabilize a queueing system whenever this is possible. In other words, dynamic instability phenomena, such as the ones reported in \cite{KS90,RS92}, are guaranteed not to occur under Max-Weight scheduling.

\par The impact of heavy-tailed traffic has been analyzed extensively in the context of single and multi-server queues; see the survey papers \cite{BBNZ03,BZ07} and the references therein. However, there is little work on more complex queueing systems and scheduling policies. A notable exception is the work by Borst \emph{et al.} \cite{BMU03}, which studies the Generalized Processor Sharing policy in the presence of heavy-tailed traffic. Related results, which come closer to the subject of this paper, are also given in \cite{MMT09}, which reports a nontrivial delay stability region for a system of parallel queues served by a single server under the Round-Robin policy. In contrast to the aforementioned scheduling policies, Max-Weight is a queue length-based policy, whose dynamics are more complex and harder to analyze. Finally, the work by Jagannathan \emph{et al.} \cite{JMMT10} presents a queueing system with heavy-tailed traffic, and under the Max-Weight policy, with a nontrivial delay stability region. However, what affects delay stability in that study is the intermittent access of servers to queues, arising, e.g., in wireless networks. Consequently, the underlying mechanism that determines the delay stability region there is very different than the one analyzed here.

\par The main contributions of this paper are as follows:
\begin{enumerate}
	\item[(i)] we show that under the Max-Weight policy, heavy-tailed arrivals at one queue may or may not cause delay instability at other queues, as determined by an intricate instability propagation phenomenon;

	\item[(ii)] we use ``fluid-type'' sample path arguments, combined with renewal theory, to prove delay instability if the arrival rate to a specific queue is greater than a certain threshold (Proposition 2);

	\item[(iii)] we use drift analysis (over a sufficiently long time interval) of a suitably constructed piecewise linear Lyapunov function to prove delay stability if the arrival rate to a specific queue is less than  that same threshold  (Proposition 3);

	\item[(iv)] as an intermediate step in establishing delay stability, we show that the expected workload of a stable M/GI/1 queue with heavy-tailed service times scales sublinearly in time, assuming just the existence of the $1+\gamma$ moment of service times (see the proof of Proposition 3).
\end{enumerate}

\par The rest of the paper is organized as follows. In Section 2 we give a detailed description of the queueing system under consideration and the probabilistic assumptions that we make, together with some necessary definitions and notation. In Sections 3 and 4 we prove our delay instability and stability results, respectively. We conclude in Section 5 with a brief discussion and future directions of research.


\bigskip\section{Model and Definitions}

\par\noindent 
Throughout the paper we denote by $\mathbb{Z}_+$ and $\mathbb{N}$ the sets of nonnegative and positive integers, respectively. Similarly, the Cartesian product of $M$ copies of $\mathbb{Z}_+$ is denoted by $\mathbb{Z}_+^M$. We use $1_E$ for the indicator variable of event $E$. Finally, $[x]^+$ stands for $\max\{x,0\}$, the nonnegative part of a scalar $x$. With few exceptions, we follow the usual convention of using lower case letters to denote real numbers or vectors and upper case letters to denote random variables or events.

\subsection{The basic model and dynamics}
\par We consider the queueing system depicted in Figure 1, consisting of three single-class, single-server queues, with infinite capacity. Time is slotted and traffic consists of constant size packets. We define 
$A_i(t)$ as the (nonnegative) number of packets (a random variable) that queue $i\in\{1,2,3\}$  receives at slot $t \in \mathbb{Z}_+$, and let $A(t)=(A_1(t),A_2(t),A_3(t))$. We assume that arrivals occur at the end of the time slots.  We define $Q_i(t)$ as the number of packets in queue $i$ at the beginning of time slot $t$, and let 
$Q(t)=(Q_1(t),Q_2(t),Q_3(t))$.

We define $S_i(t)$ as the number of packets that the servers attempt to remove from queue $i$ during time slot $t$, and let $S(t)=(S_1(t),S_2(t),S_3(t))$.
We assume that $S_i(t)$ can only take values in $\{0,1\}$. 
An attempt will be successful if and only if queue $i$ has a packet, so that the number of packets actually removed from queue $i$ during time slot $t$ is $S_i(t) \cdot 1_{\{Q_i(t)>0\}}$. Accordingly, 
the \textbf{queue length dynamics} take the form
\begin{equation}
Q_i(t+1) = Q_i(t) + A_i(t) - S_i(t) \cdot 1_{\{Q_i(t)>0\}}, \qquad t \in \mathbb{Z}_+, \qquad i \in \{1,2,3\},\nonumber
\end{equation}
initialized with an arbitrary vector of initial queue lengths $Q(0)$ in $\mathbb{Z}_+^3$.

\begin{figure}[ht]
\centering
\includegraphics[scale=0.5]{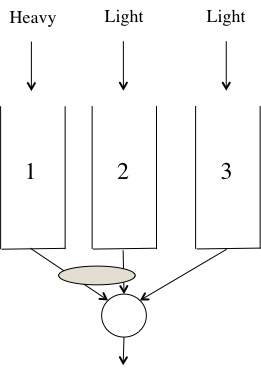}
\caption{A simple system of three single-class, single-server queues. A scheduler decides, at each time slot, whether queues 1 and 2 will be served simultaneously, or whether queue 3 will be served alone. Queue 1 receives heavy-tailed traffic, whereas queues 2 and 3 receive light-tailed traffic.}\label{fig:Fig1}
\end{figure}

\subsection{Scheduling constraints and the Max-Weight policy}
We assume that the servers of queues 1 and 2 can be active simultaneously, whereas the server of queue 3 can only be active alone. ``Scheduling constraints'' of this type are common in communication networks, such as wireless networks or data switches. Formally, the sets of queues $\{1,2\}$ and $\{3\}$, as well as the empty set, are called the \textbf{feasible schedules}. Furthermore, $S(t)$ is constrained to take values in the set $\mathcal{S}=\Big\{ (0,0,0),(1,1,0),(0,0,1) \Big\}$, for all $t \in \mathbb{Z}_+$. 
The service discipline of packets within each queue is assumed to be ``\textbf{First Come, First Served}.'' 

The choice of a service vector $S(t)\in\mathcal{S}$ is made at each time slot according to a scheduling policy. In this paper, we focus exclusively on the Max-Weight policy, which chooses a schedule with the maximum total workload, at each time slot. Formally, 
\begin{equation}
S(t) \ \in \ \arg\max_{S \in \mathcal{S}} \Big\{ \sum_{i=1}^3 Q_i(t) \cdot S_i \Big\}, \qquad t \in \mathbb{Z_+}. \nonumber
\end{equation}
In case of a tie, each one of the feasible schedules is chosen with equal probability.

\subsection{Assumptions on the arrival processes}
We assume that the discrete time stochastic arrival processes $\{A_i(t);\ t \in \mathbb{Z}_+\},\ i \in \{1,2,3\}$ are independent and identically distributed (IID) over time, and  mutually independent. We denote by $\lambda_i = E[A_i(0)]$ the arrival rate of the $i^{th}$ process, and let $\lambda=(\lambda_1,\lambda_2,\lambda_3)$  For the model to be interesting, all rates are assumed to be strictly positive. 

We say that a random variable $X$ is {\bf heavy-tailed} (respectively, {\bf light-tailed}) if $\mathbb{E}[X^2]$ is infinite (respectively, finite). Similarly, we say that the $i^{th}$ arrival process is heavy or light-tailed depending on whether $\E[A_i^2(t)]$ is infinite or finite, respectively. Throughout the paper we assume that the first arrival process is heavy-tailed, and the other two arrival processes are light-tailed.

\subsection{The stability region}
A batch of packets arriving to a queue at any given time slot can be viewed as a single entity, and will be referred to as a {\bf file}.  We define the \textbf{end-to-end delay of a file} to be the number of time slots that the file spends in the system, starting from the time slot right after it arrives, until the time slot that its last packet exits the system. (For example, a two-packet file that arrives during time slot $t$ and is served during the next two time slots has an end-to-end delay of 2.)
For $k \in \mathbb{N}$, we denote by $D_i(k)$ the end-to-end delay of the $k^{th}$ file arriving at queue $i$, and  let $D(k)=(D_1(k),D_2(k),D_3(k))$.

The following definition gives the precise notion of stability that we use in this paper.

\medskip\par\textbf{Definition 2: (Stability)} The earlier defined queueing system  is stable under a particular scheduling policy if the vector-valued sequences $\{Q(t);\ t \in \mathbb{Z}_+\}$ and $\{D(k);\ k \in \mathbb{N}\}$ converge in distribution, and their limiting distributions do not depend on $Q(0)$.

\medskip\par Notice that our definition of stability is slightly different than the commonly used one (positive recurrence of the Markov chain of queue lengths), because it includes the convergence of the sequence of file delays $\{D(k);\ k \in \mathbb{N}\}$. The reason is that we are interested in properties of the limiting distribution of $\{D(k);\ k \in \mathbb{N}\}$ and, naturally, we need to ensure that this limiting distribution exists.

An arrival rate vector $\lambda$ is said to belong to the \textbf{stability region} of the system, if there exist nonnegative real numbers $\mu_{12}$ and  $\mu_3$, such that
\begin{align}
\max\{\lambda_1, \lambda_2\} &\leq \mu_{12}, \nonumber \\
\lambda_3 &\leq \mu_3, \nonumber \\
\mu_{12} + \mu_3 &< 1. \nonumber
\end{align}
Intuitively, one can think of $\mu_{12}$ and $\mu_3$ as the fractions of time that the schedules $(1,1,0)$ and $(0,0,1)$, respectively, are applied. If a positive arrival vector $\lambda$ is outside (and at a positive distance) from the stability region, it is easily shown that there is no scheduling policy that makes the system stable. In contrast, when $\lambda$ belongs to the stability region, the system can be made stable, e.g., by using the Max-Weight policy; see Lemma 1 below. Because of these considerations, we will assume throughout the rest of the paper that $\lambda$ belongs to the stability region.

\medskip\par\textbf{Lemma 1: (Stability under Max-Weight)} The queueing system under consideration is stable under the Max-Weight scheduling policy, for all positive arrival rate vectors in the stability region.

\begin{proof}
It can be verified that the sequence $\{Q(t);\ t \in \mathbb{Z}_+\}$ is a time-homogeneous, irreducible, and aperiodic Markov chain on the countable state-space $\mathbb{Z}_+^3$. Proposition 2 of \cite{S04} implies that this Markov chain is also positive recurrent. Hence, $\{Q(t);\ t \in \mathbb{Z}_+\}$ converges in distribution, and its limiting distribution does not depend on $Q(0)$. Based on this, it can be verified that the sequence $\{D(k);\ k \in \mathbb{N}\}$ is a (possibly delayed) aperiodic and positive recurrent regenerative process. Therefore, it also converges in distribution, and its limiting distribution does not depend on $Q(0)$; see \cite{SW93}.
\end{proof}

\medskip\par 
We let $Q=(Q_1,Q_2,Q_3)$ and $D=(D_1,D_2,D_3)$ be random vectors distributed according to the limiting distributions of $\{Q(t);\ t \in \mathbb{Z}_+\}$ and $\{D(k);\ k \in \mathbb{N}\}$, respectively, under the Max-Weight scheduling policy, and refer to them as the steady-state queue lengths and delays.

\subsection{Delay stability}
We now define formally, and then discuss, the property that we will be focusing on.

\medskip\par\textbf{Definition 3: (Delay Stability)} Queue $i \in \{1,2,3\}$ is delay stable under a particular scheduling policy if the system  is stable under that policy, and $\mathbb{E}[D_i]$ is finite; otherwise, queue $i$ is delay unstable.

\medskip\par The following lemma relates the steady-state quantities $\mathbb{E}[Q_i]$ and $\mathbb{E}[D_i]$, and will help us prove delay stability results.

\medskip\par\textbf{Lemma 2 \cite{MMTpre}:} For the system under consideration, and under Max-Weight scheduling,
\begin{equation}
\mathbb{E}[Q_i]<\infty \ \Longleftrightarrow \ \mathbb{E}[D_i]<\infty, \qquad i \in \{1,2,3\}. \nonumber
\end{equation}

\medskip In the remainder  of the paper we will focus on the delay stability of queue 2. This is because under the Max-Weight scheduling policy, queues 1 and 3 are always delay unstable, as established formally in Theorems 1 and 2 of the companion paper \cite{MMTpre}.

\medskip\par\textbf{Proposition 1 \cite{MMTpre}:} For the system under consideration, and under Max-Weight scheduling, queues 1 and 3 are delay unstable, for all positive arrival rate vectors in the stability region.

\medskip
We provide some intuition as to why Proposition 1 is true. For queue 1, the result follows easily from the Pollaczek-Khinchine formula for the expected delay in a stable M/GI/1 queue, and a stochastic comparison argument. Regarding queue 3, we argue as follows. Queue 1 is occasionally very long (infinite, in steady-state expectation), due to the heavy-tailed nature of the traffic that it receives. Whenever queue 1 becomes very long, queue 3 is starved and builds up, at a more or less constant rate, to a size comparable to that of queue 1. Thus, large values of $Q_1(t)$ result in large values of $Q_3(t)$ at subsequent times. As $\E[Q_1]$ is infinite, we can argue that $\E[Q_3]$ is infinite as well, and Lemma 2 implies that queue 3 is delay unstable.

\medskip Our main results will concern queue 2. We will show in the next two sections that it is delay unstable (respectively, delay stable) if $\lambda_2$ is greater (respectively, less) than $(1+\lambda_1-\lambda_3)/2$.


\bigskip\section{Delay Instability Region}

\par Lemma 1 and Proposition 1 in the previous section establish that the queueing system under consideration is stable, and that queues 1 and 3 are delay unstable for every positive arrival rate vector in the stability region. Regarding queue 2, the situation is more complicated. One could argue that queue 1, by virtue of its receiving heavy-tailed traffic, would tend to claim service more often, and that queue 2 would benefit from the resulting additional servive opportunities. Somewhat surprisingly though, there is a more intricate indirect mechanism at play that can render queue 2 delay unstable.

\medskip\par\textbf{Proposition 2:} For the system under consideration, with positive arrival rates within the stability region, and under Max-Weight scheduling, if the arrival rates satisfy $\lambda_2>(1+\lambda_1-\lambda_3)/2$, then queue 2 is delay unstable.

\medskip\par  Before proceeding with the formal proof of Proposition 2, we provide an intuitive outline of the argument, also aimed at explaining the threshold value $(1+\lambda_1-\lambda_3)/2$. Our approach is based on tracking the evolution of the system on a particular set of ``fluid'' sample paths:\footnote{In essence, our argument employs a ``fluid approximation''; however, instead of establishing formal convergence to a fluid model (which would be unduly complicated for our purposes), we proceed in an elementary manner, arguing from first principles.} assume that at time slot 0, queue 1 receives a very large file, consisting of $b$ packets. For a long period of time after that, queue 3 does not receive service under the Max-Weight policy, and builds up. If the arrival processes of all traffic flows are close to their ``average behavior'' (in the Strong Law of Large Numbers sense), then at the time slot when the service switches from schedule $\{1,2\}$ to schedule $\{3\}$, the lengths of both queues 1 and 3 are proportional to $b$, whereas queue 2 is still small. From that point on, the Max-Weight policy will drain the weights of the two schedules at roughly the same rate, until one of the weights becomes zero. 

Let $\mu_i$ be the average departure rate from queue $i$ during the latter  period. For the weights of the two schedules to be drained at the same rate, the departure rates have to satisfy:
\begin{equation}
\lambda_1 + \lambda_2 - \mu_1 - \mu_2 = \lambda_3 - \mu_3. \nonumber
\end{equation}
Moreover, the fact that Max-Weight is a work-conserving policy implies that
\begin{equation}
\mu_1 + \mu_3 = 1. \nonumber
\end{equation}
Finally, since queues 1 and 2 are served simultaneously, and queue 2 may be empty through parts of the draining period, we have that
\begin{equation}
\mu_1 \geq \mu_2. \nonumber
\end{equation}
The above equations and some simple algebra imply that
\begin{equation}
\mu_2 \leq \frac{1+\lambda_1+\lambda_2-\lambda_3}{3}. \nonumber
\end{equation}

Suppose that the arrival rates satisfy
$$
\lambda_2 > \frac{1+\lambda_1-\lambda_3}{2}.$$
Then, 
$$\lambda_2 > \frac{1+\lambda_1+\lambda_2-\lambda_3}{3} \geq \mu_2.$$
This implies that queue 2 builds up at a roughly constant rate, to size $O(b)$, during a period of time whose duration is proportional to $b$, and the integral of $Q_2$ over a busy period of the process becomes of order $O(b^2)$. Because $b$ is drawn from a heavy-tailed distribution (infinite second moment), it follows that $\mathbb{E}[Q_2]$ is infinite.

\begin{proof}
\par We start by defining a shorthand notation that we will be using in the course of the proof. We say that a random variable $X$ {\it scales at least linearly} with $b$ on the event $H$, and write $X=\Omega_H(b)$, if there exist positive constants $k$ and $k'$ (possibly depending on the event $H$), such that $X \geq k \cdot b - k'$, for all sample paths in $H$.

\par We break the proof into four steps, which follow the various stages in our earlier proof outline.

\medskip{{\bf Step 1: buildup of queue 3 following a large arrival to queue 1.}}
\par\noindent
Because the system is stable, it empties infinitely often, and the times at which $Q(t)=0$ are renewal epochs. Let us consider the system at a typical renewal epoch which, for simplicity of notation, we assume to happen at time 0. Consider the set of sample paths for which, at time slot 0, queue 1 receives a file that consists of $b$ packets, and all other queues receive no traffic; we denote this set of sample paths by $H(b)$. Since $\lambda_i<1$ (stability), and since $1-\P(A_i(0)=0) =\P(A_i(0)>0) \leq \E[A_i(0)]=\lambda_i<1$, we have $\P(A_i(0)=0)>0$, for all $i$. Let $\B$ be the support of the distribution of $A_1(0)$. Using the independence of the arrival processes, we have, for every $b\in\B$,
\begin{equation}
\mathbb{P}(H(b))= \mathbb{P}(A_1(0)=b) \cdot \mathbb{P}(A_2(0)=0) \cdot \mathbb{P}(A_3(0)=0) >0. \nonumber
\end{equation}

\par For sample paths in $H(b)$, we denote by $T^1_b$ the first time slot, starting from 0, when the length of queue 3 becomes greater than or equal to the sum of the lengths of queues 1 and 2:
\begin{equation}
T^1_b = \min \{t>0 \mid Q_3(t) \geq Q_1(t)+Q_2(t) \} \cdot 1_{H(b)}. \nonumber
\end{equation}

\noindent The first part of the proof is to show that $Q_1(T^1_b)$ and $Q_3(T^1_b)$ scale at least linearly with $b$, provided all arrival processes are close to their ``average behavior.''

\par By the definition of the stopping time $T^1_b$, we have
\begin{equation}
Q_1(T^1_b) \leq Q_1(T^1_b) + Q_2(T^1_b) \leq Q_3(T^1_b). \nonumber
\end{equation}
A direct consequence of the Strong Law of Large Numbers is that for every $\epsilon>0$ there exists $\delta>0$, such that the set of sample paths
\begin{equation}
\Delta = \Big\{ (\lambda_i-\epsilon)t-\delta \leq \sum_{\tau=1}^{t} A_i(\tau) \leq (\lambda_i+\epsilon)t+\delta, \ \ \forall t \in \mathbb{N}, \  \ \forall i \in \{1,2,3\} \Big\}, \nonumber
\end{equation}
has positive probability. Consequently, the set of sample paths
\begin{equation}
\Delta(b) = \Big\{ (\lambda_i-\epsilon)t-\delta \leq \sum_{\tau=1}^{t} A_i(\tau) \leq (\lambda_i+\epsilon)t+\delta, \ \ \ \forall t \in \{1,\ldots,T_b^1-1\}, \ \forall i \in \{1,2,3\} \Big\}, \nonumber
\end{equation}
has positive probability, uniformly over all $b$.

From now on we fix a small $\epsilon>0$. (How small it will have to be will become apparent in the course of the proof.) We then fix a corresponding $\delta$ such that $\inf_b \P(\Delta(b)) \geq \P(\Delta) >0$. Let
$\tilde{H}(b)$ be the set of sample paths $H(b) \cap \Delta(b)$, and observe that $H(b) \cap \Delta(b) \supset H(b) \cap \Delta$. Then, the IID nature of the arriving traffic implies that $H(b)$ and $\Delta$ are independent, so that
$$
\mathbb{P}(\tilde{H}(b))=\mathbb{P}(H(b) \cap \Delta(b)) \geq \mathbb{P}(H(b) \cap \Delta) =  \mathbb{P}(H(b)) \cdot P(\Delta)>0. 
$$

\par Recall that at most one packet can be removed from each queue at any given time slot. So, for sample paths in $\tilde{H}(b)$, we have that
\begin{equation}
Q_1(T^1_b) \geq b -(T^1_b-1) + (\lambda_1-\epsilon) \cdot (T^1_b-1) -\delta. \nonumber
\end{equation}
Moreover, queue 3 receives no service before time slot $T^1_b$ under the Max-Weight scheduling policy, which implies that
\begin{equation}
Q_3(T^1_b) = \sum_{t=1}^{T^1_b-1} A_3(t) \leq (\lambda_3+\epsilon) \cdot (T^1_b-1) + \delta. \nonumber
\end{equation}
Since $Q_1(T^1_b) \leq Q_3(T^1_b)$, the last two inequalities and some algebra yield
\begin{equation}
T^1_b - 1 \geq \frac{b - 2 \delta}{1+\lambda_3-\lambda_1 + 2 \epsilon}. \tag{1}
\end{equation}
(This argument requires the last denominator to be positive. This will be the case as long as $\epsilon$ has been chosen small enough.)
Therefore,
\begin{align}
Q_3(T^1_b) = \sum_{t=1}^{T^1_b-1} A_3(t) \geq (\lambda_3-\epsilon) \cdot (T^1_b-1) - \delta \geq (\lambda_3-\epsilon) \cdot \frac{b - 2 \delta}{1+\lambda_3-\lambda_1 + 2 \epsilon} - \delta, \tag{2}
\end{align}
which implies that $Q_3(T^1_b)=\Omega_{\tilde{H}(b)}(b)$, since we can chose $\epsilon$ to be less than $\lambda_3$.

Coming to queue 2, it can be verified that for sample paths in $\Delta$ (and hence in $\tilde{H}(b)$), and for any subinterval $\{\tau_0,\ldots,\tau_1\}$ of $\{1,\ldots,T^1_b\}$,
\begin{equation} 
\sum_{t=\tau_0}^{\tau_1-1} A_2(t) \leq (\lambda_2+\epsilon) \cdot (\tau_1-\tau_0) +2 \epsilon \tau_0 + 2 \delta. \nonumber
\end{equation} 
We assume that $\epsilon$ has been chosen so that $\lambda_2+\epsilon<1$. Recall also that queue 2 gets served whenever it is nonempty throughout the period $\{1,\ldots,T^1_b-1\}$. We use Lindley's formula, the above upper bound on the arrivals to queue 2, and the fact that since queue 2 gets served whenever it is nonempty throughout the period $\{1,\ldots,T^1_b-1\}$, to conclude that
\begin{equation}
Q_2(T^1_b) \leq A_2(T^1_b-1) + 2 \epsilon (T^1_b-1) + 2 \delta \leq \lambda_2 + 4 \epsilon (T^1_b-1) + 4 \delta. \tag{3}
\end{equation}
This shows that, essentially, $Q_2(T^1_b)$ does not scale with $b$.

\par We finally turn our attention to $Q_1(T^1_b)$. By definition of the stopping time $T^1_b$,
\begin{equation}
Q_3(T^1_b-1) < Q_1(T^1_b-1)+Q_2(T^1_b-1). \tag{4}
\end{equation}
By arguing similar to the derivation of Eqs.\ (2) and (3), it can be verified that
\begin{equation}
Q_2(T^1_b-1) \leq \lambda_2 + 4 \epsilon (T^1_b-2) + 4 \delta, \tag{5}
\end{equation}
and that $Q_3(T^1_b-1)=\Omega_{\tilde{H}(b)}(b)$. Then, Eqs.\ (4) and (5) readily imply that $Q_1(T^1_b)=\Omega_{\tilde{H}(b)}(b)$, when  $\epsilon$ is chosen sufficiently small.

To summarize,  at time slot $T^1_b$, and for sample paths in $\tilde{H}(b)$, the lengths of queues 1 and 3 are proportional to $b$, while queue 2 is still small.

\medskip{{\bf Step 2: draining down until queue 1 or 3 empties.}}

\par\noindent Let $T^2_b$ be the first time slot after $T_b^1$ that either queue 1 or queue 3 becomes empty:
\begin{equation}
T^2_b = \min \{t>T_b^1 \mid Q_1(t) \cdot Q_3(t)=0 \} \cdot 1_{\tilde{H}(b)}. \nonumber
\end{equation}
We will show that if the arrival processes stay close to their ``average behavior'' (i.e., the event $\tilde{H}(b)$ occurs), then, at time slot $T^2_b$, the length of queue 3 cannot be much larger than the sum of the lengths of queues 1 and 2. The reason is that $Q_1(t)+Q_2(t)$ and $Q_3(t)$ are kept roughly equal by the Max-Weight policy throughout the interval $[T_b^1,T_b^2]$.

\par For the same constants $\epsilon$ and $\delta$ as Step 1, the set of sample paths
\begin{equation}
\Delta'(b) = \Big\{ (\lambda_i-\epsilon)t-\delta \leq \sum_{\tau=T_b^1}^{t} A_i(\tau) \leq (\lambda_i+\epsilon)t+\delta, \ \forall t \in \{T_b^1,\ldots,T_b^2-1\}, \ \forall i \in \{1,2,3\} \Big\} \nonumber
\end{equation}
contains $\Delta$. Let $\hat{H}(b) = \tilde{H}(b) \cap \Delta'(b)$. The events $\Delta(b)$ and $\Delta'(b)$ are determined by the arrivals over disjoint time intervals. Hence, due to the IID nature of the arrival processes, 
$\Delta(b)$ and $\Delta'(b)$ are independent. Since they both contain the positive probability event $\Delta$, we have
\begin{equation}
\mathbb{P}(\hat{H}(b)) \geq \mathbb{P}(H(b)) \cdot \big( \mathbb{P}(\Delta) \big)^2>0. \nonumber
\end{equation}
We will show that for sample paths in $\hat{H}(b)$,
\begin{equation}
Q_3(T^2_b) \leq Q_1(T^2_b) + Q_2(T^2_b) + 2 \epsilon (T^2_b-T^1_b) + 2 \delta + 3. \tag{6}
\end{equation}

We first notice that queues 1 and 3 cannot empty at the same time slot, since they cannot be served simultaneously. Therefore, we have two possible cases: either $Q_3(T^2_b)=0$, in which case Eq. (6) is trivially satisfied, or  $Q_1(T^2_b)=0$, which we henceforth assume. In the latter case, $S_1(T^2_b-1)=S_2(T^2_b-1)=1$, and $S_3(T^2_b-1)=0$. For sample paths in $\hat{H}(b)$, we have that
\begin{align}
Q_3(T^2_b) = Q_3(T^2_b-1) + A_3(T^2_b-1) \leq Q_3(T^2_b-1) + \lambda_3 + 2 \epsilon \cdot (T^2_b-T^1_b) + 2 \delta. \tag{7}
\end{align}
Moreover, under the Max-Weight scheduling policy, and in order for the for the set of queues $\{1,2\}$ to be served at time slot $T^2_b-1$,
\begin{equation}
Q_3(T^2_b-1) \leq Q_1(T^2_b-1) + Q_2(T^2_b-1). \tag{8}
\end{equation}

\noindent Finally,
\begin{equation}
Q_1(T^2_b-1) + Q_2(T^2_b-1) - 2 \leq Q_1(T^2_b) + Q_2(T^2_b). \tag{9}
\end{equation}
Eq.\ (6) follows immediately, by combining Eqs.\ (7)-(9) and using also the fact $\lambda_3<1$.

\medskip
{\bf Step 3: growth of queue 2.}
\par\noindent
At time $T^2_b$, $Q_3(T^2_b)$ cannot much larger than $Q_1(T^2_b)+Q_2(T^2_b)$. Therefore, queue 3 must have been receiving a certain fraction of the total service between times $T^1_b$ and $T^2_b$. We will show that this results in queue 2 not receiving enough service, and that $Q_2$ starts growing. In particular,  if $\lambda_2>(1+\lambda_1-\lambda_3)/2$, then, for the  sample paths of interest, $Q_2(T^2_b)=\Omega_{\hat{H}(b)}(b)$.

By definition,
\begin{equation}
Q_3(T^1_b) \geq Q_1(T^1_b) + Q_2(T^1_b). \nonumber
\end{equation}
By subtracting the two sides of this inequality from Eq. (6), we get
\begin{equation}
Q_3(T^2_b)-Q_3(T^1_b) \leq Q_1(T^2_b)-Q_1(T^1_b) + Q_2(T^2_b)-Q_2(T^1_b) + 2 \epsilon (T^2_b-T^1_b) + 2 \delta+3. \tag{10}
\end{equation}

\par For sample paths in $\hat{H}(b)$, define the random variables
\begin{equation}
\mu_i = \Big( \frac{1}{T^2_b-T^1_b} \cdot \sum_{t=T^1_b}^{T^2_b-1} S_i(t) \Big) \cdot 1_{\hat{H}(b)}, \qquad i \in \{1,2,3\}, \nonumber
\end{equation}

\noindent which are the average service rates to each queue during the interval $\{T^1_b,\ldots,T^2_b-1\}$. Notice that $\mu_1=\mu_2$ and $\mu_1+\mu_3=1.$

\par Since both queues 1 and 3 are nonempty during the inerval $\{T^1_b,\ldots,T^2_b-1\}$, we have
\begin{align}
Q_1(T^2_b)-Q_1(T^1_b) &\leq (\lambda_1+\epsilon-\mu_1) \cdot (T^2_b-T^1_b)+\delta, \tag{11} \\
Q_3(T^2_b)-Q_3(T^1_b) &\geq (\lambda_3-\epsilon-\mu_3) \cdot (T^2_b-T^1_b)-\delta. \tag{12}
\end{align}
Eqs. (10), (11), and (12) imply that
\begin{align}
(\lambda_3-\epsilon-\mu_3) \cdot (T^2_b-T^1_b)-\delta \leq &(\lambda_1+\epsilon-\mu_1) \cdot (T^2_b-T^1_b)+\delta \nonumber \\
&+ Q_2(T^2_b)-Q_2(T^1_b) + 2 \epsilon (T^2_b-T^1_b) + 2 \delta+3. \nonumber
\end{align}

\noindent We replace  $\mu_3$ by $1-\mu_1$ and collect terms, to obtain
\begin{align}
-\mu_1 \cdot (T^2_b-T^1_b) &\geq - \Big( \frac{1+\lambda_1-\lambda_3+4 \epsilon}{2} \Big) \cdot (T^2_b-T^1_b) + \frac{Q_2(T^1_b)-Q_2(T^2_b)}{2} - \frac{4 \delta +3}{2} \nonumber \\
&\geq - \Big( \frac{1+\lambda_1-\lambda_3+ 4 \epsilon}{2} \Big) \cdot (T^2_b-T^1_b) - \frac{Q_2(T^2_b)}{2} - \frac{4 \delta+3}{2}. \nonumber
\end{align}
For  sample paths in $\hat{H}(b)$, we use the definition of $\Delta'(b)$ to upper bound the number of arrivals to queue 2. We also use the fact that queue 2 has $\mu_2(T_b^2-T_b^1)$  service opportunities, with $\mu_2=\mu_1$, and obtain
\begin{align}
Q_2(T^2_b) &\geq (\lambda_2 - \epsilon -\mu_1) \cdot (T^2_b-T^1_b) -\delta \nonumber \\
&\geq \Big( \lambda_2 - \frac{1+\lambda_1-\lambda_3}{2} - 3 \epsilon \Big) \cdot (T^2_b-T^1_b) -\frac{Q_2(T^2_b)}{2}-\frac{6 \delta+3}{2}. \nonumber
\end{align}
Therefore,
\begin{equation}
Q_2(T^2_b) \geq \frac{2}{3} \cdot \Big( \lambda_2 - \frac{1+\lambda_1-\lambda_3}{2} - 3 \epsilon \Big) \cdot (T^2_b-T^1_b) -2 \delta - 1. \nonumber
\end{equation}

\noindent If $\lambda_2>(1+\lambda_1-\lambda_3)/2$, the constant $\epsilon$ can be chosen sufficiently small so that
\begin{equation}
\lambda_2 - \frac{1+\lambda_1-\lambda_3}{2} - 3 \epsilon>0. \nonumber
\end{equation}

\par A final observation is that the duration of the interval $\{T^1_b,\ldots,T^2_b-1\}$ is bounded from below by $\min\{Q_1(T^1_b),Q_3(T^1_b)\}$, because both queues are served at unit rate. Therefore, $T^2_b-T^1_b=\Omega_{\tilde{H}(b)}(b)$ and
\begin{equation}
Q_2(T^2_b)=\Omega_{\hat{H}(b)}(b). \tag{13}
\end{equation}

\medskip
{\bf Step 4: the growth scenario for queue 2 implies large average queue size.}

\par In Step 3 we showed that for sample paths in $\hat{H}(b)$, queue 2 builds up to the order of $b$. We use this fact, and renewal theory, to show that the steady-state expected length of queue 2 is infinite. The sequence of times at which $Q(t)=0$ are renewal epochs. Denote by $X_i$ the length of the $i^{th}$ inter-renewal period. The random variables $\{X_i; \ i \in \mathbb{N}\}$ can be viewed as IID copies of some nonnegative random variable $X$, with finite first moment; this is because the empty state is positive recurrent under the Max-Weight policy (see Proposition 2 of \cite{S04}).

\par We define an instantaneous reward on this renewal process:
\begin{equation}
R^M(t) = \min\{Q_2(t),M\}, \qquad \ t \in \mathbb{Z}_+, \nonumber
\end{equation}
\noindent where $M$ is a positive integer. 

\par Eq.\ (13) implies that there exist positive constants $c$ and $b_0$, such that
\begin{equation}
Q_2(T^2_b) \geq cb, \qquad \forall\ b \geq b_0, \nonumber
\end{equation}
for all sample paths in $\hat{H}(b)$.

\par Since at most one packet from queue 2 can be served at each time slot, the length of queue 2 is at least $cb/2$ packets over a time period of length at least $cb/2$ time slots. Hence, the aggregate reward $R^M_{agg}$, i.e., the reward accumulated over a renewal period, satisfies the lower bound
\begin{equation}
R^M_{agg} \cdot 1_{\{b \geq b_0\}} \cdot 1_{\hat{H}(b)} \geq \min \Big\{ \Big( \frac{cb}{2}\Big)^2 \cdot 1_{\{b \geq b_0\}}, M^2 \Big\} \cdot 1_{\hat{H}(b)}. \nonumber
\end{equation}
Then, the expected aggregate reward is bounded below by
\begin{align}
\mathbb{E}[R^M_{agg}] \geq \big(\mathbb{P}(\Delta)\big)^2 \cdot \mathbb{P}(A_2(0)=0) \cdot \mathbb{P}(A_3(0)=0) \cdot \sum_{b=1}^{\infty} \min \Big\{ \Big( \frac{cb}{2}\Big)^2 \cdot 1_{\{b \geq b_0\}}, M^2 \Big\} \cdot \mathbb{P}(A_1(0)=b). \nonumber
\end{align}

\noindent So, there exists a positive constant $c'$ such that
\begin{equation}
c' \cdot \mathbb{E} \Big[ \min \Big\{ \Big( \frac{c A_1(0)}{2}\Big)^2 \cdot 1_{\{A_1(0) \geq b_0\}}, M^2 \Big\} \Big] \leq \mathbb{E}[R^M_{agg}]. \tag{14}
\end{equation}

\par We have argued that inter-renewal periods have finite expectation, and, clearly, the expected aggregate reward is finite. Then, the Renewal Reward theorem (e.g., see Section 3.4 of \cite{G96}) implies that
\begin{equation}
\frac{\mathbb{E}[R^M_{agg}]}{\mathbb{E}[X]} = \lim_{T \to \infty} \frac{1}{T} \sum_{t=0}^{T-1} R^M(t), \qquad \mbox{w.p.1.} \tag{15}
\end{equation}

\noindent Also, the fact that the reward is a bounded function of an ergodic Markov chain implies that
\begin{equation}
\lim_{T \to \infty} \frac{1}{T} \sum_{t=0}^{T-1} R^M(t) = \mathbb{E}[\min\{Q_2,M\}], \qquad \mbox{w.p.1.} \tag{16}
\end{equation}

\noindent Eqs. (14)-(16) imply that
\begin{equation}
\frac{c'}{\mathbb{E}[X]} \cdot \mathbb{E} \Big[ \min \Big\{ \Big(\frac{c A_1(0)}{2}\Big)^2 \cdot 1_{\{ A_1(0) \geq b_0\}} , M^2 \Big\} \Big] \leq \mathbb{E}[\min\{Q_2,M\}]. \nonumber
\end{equation}

\noindent Taking the limit as $M$ goes to infinity on both sides, and using the Monotone Convergence theorem (e.g., see Section 5.3 of \cite{WI91}) gives
\begin{equation}
\frac{c' \cdot c^2}{4 \mathbb{E}[X]} \cdot \mathbb{E}[A_1^2(0) \cdot 1_{\{A_1(0) \geq b_0\}}] \leq \mathbb{E}[Q_2]. \nonumber
\end{equation}

\noindent Finally, the fact that the random variable $A_1(0)$ is heavy-tailed, implies that $\mathbb{E}[Q_2]$ is infinite. Combined with Lemma 2,  this gives the desired result.
\end{proof}


\bigskip\section{Delay Stability Region}

In this section we establish that when $\lambda_2<(1+\lambda_1-\lambda_3)/2$, then queue 2 is delay stable. Notice that this result involves the same threshold as in Proposition 2, and we therefore have an exact characterization of the delay stability region.

The proof of this result relies on drift analysis of a suitable Lyapunov function. The usual analysis of the Max-Weight policy involves a quadratic Lyapunov function \cite{TE92}. Different types of Lyapunov functions have also been used for stability and performance analysis of other queueing networks and policies, e.g., piecewise linear functions \cite{DM97,BGT01}, and norms \cite{STZ10,VLYS10}. However, in the presence of heavy-tailed traffic, the expected value of all these Lyapunov functions in steady-state is infinite, which renders their drift analysis uninformative.

Here we consider a carefully constructed piecewise linear Lyapunov function that is \textbf{not} radially unbounded (i.e., the value of the Lyapunov function need not be large when the queue lengths are large), and which has a negative drift only when $\lambda_2<(1+\lambda_1-\lambda_3)/2$. A common technical difficulty with piecewise linear Lyapunov functions is that the stochastic descent property is often lost at locations where the Lyapunov function is nondifferentiable. This difficulty can be handled by either smoothing the Lyapunov  function (e.g., as in \cite{DM97}), or by showing that the stochastic descent property still holds if we look ahead a sufficiently large number of time slots (e.g., as in \cite{T87}). We follow the second approach.

Furthermore, we also assume that light-tailed traffic has exponentially decaying tails, which allows us to exploit  the tools in \cite{H82}. Formally, a nonnegative random variable $X$ is \textbf{exponential-type}, if there exists some $\theta>0$ such that $\mathbb{E}[\exp(\theta X)]$ is finite.\footnote{We conjecture that the statement of Proposition 3 is true for any type of light-tailed traffic. However, the results in \cite{H82} cannot be invoked without the additional assumption of exponential-type traffic.}

\medskip\par\textbf{Proposition 3:} For the system under consideration, suppose that the arrival rate vector is positive and within the stability region, and that the Max-Weight policy is used. Suppose furthermore that (i) $A_2(0)$ and $A_3(0)$ are exponential-type, and (ii) there exists some $\gamma>0$ such that $\E[A_1^{1+\gamma}(0)]<\infty$. If the arrival rates satisfy $\lambda_2<(1+\lambda_1-\lambda_3)/2$, then queue 2 is delay stable, and the steady-state length of queue 2 is exponential-type.

\medskip\par Before we proceed with the formal proof of Proposition 3, we provide the underlying intuition by arguing (loosely) in terms of a fluid approximation. Our analysis rests on drift analysis of the following piecewise linear Lyapunov function
\begin{equation}
V(t) = 3 Q_2(t) + \Big[ Q_3(t)-Q_1(t)-Q_2(t) \Big]^+, \qquad t \in \mathbb{Z}_+. \tag{17}
\end{equation}
Our goal is to establish a negative drift for $V(t)$, which, combined with the special structure of the problem (the arrivals to queues 2 and 3 being exponential-type), will imply that
\begin{equation}
\mathbb{E} \Big[ 3 Q_2 + [ Q_3-Q_1-Q_2]^+ \Big] < \infty, \nonumber
\end{equation}
\noindent where the expectation is taken with respect to steady-state distributions. Together with Lemma 2, this will imply the delay stability of queue 2.

\par The suitability of this Lyapunov function can be seen as follows. If $Q_1(t)+Q_2(t) > Q_3(t)$, then the Lyapunov function reduces to $3 Q_2(t)$, which can be easily shown to have negative drift (as long as $Q_2>0$), because queue 2 is served under Max-Weight in that region. If, on the other hand, $Q_1(t)+Q_2(t) < Q_3(t)$, then the Lyapunov function reduces to $2 Q_2(t) + Q_3(t)-Q_1(t)$. In that region, queue 3 is served under Max-Weight, so the drift of $V(t)$ is equal to $2 \lambda_2 + \lambda_3 -\lambda_1 - 1$, which is strictly negative by our assumption on $\lambda_2$. See Figure 2 for a geometric interpretation.

\begin{figure}[ht]
\centering
\includegraphics[scale=0.5]{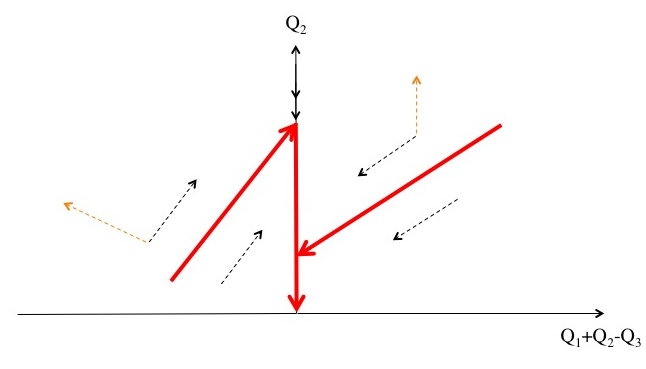}
\caption{A two-dimensional representation of the dynamics of the system when $\lambda_2<(1+\lambda_1-\lambda_3)/2$. The dashed black arrows represent the gradient of the queue length vector under the Max-Weight policy, and the dashed orange arrows represent the gradient of the Lyapunov function, in different regions of the state space. The solid red arrows represent typical trajectories of the associated fluid approximation. Our Lyapunov function has been chosen so that an obtuse angle is formed between the two in all regions of the state space, establishing a uniform negative drift over the entire state space.}\label{fig:Fig2}
\end{figure}

\par The analysis becomes subtler at the boundary between the two regions of the state space, namely when the weights, $Q_1(t)+Q_2(t)$ and $Q_3(t)$, of the two candidate schedules are equal. Once at the boundary, the Max-Weight policy keeps the state (of the fluid model) there. Similar to our discussion of Proposition 2, the resulting departure rate from queue 2 satisfies 
$$
\lambda_1 + \lambda_2 - \mu_1 - \mu_2 = \lambda_3 - \mu_3. $$
Using also the properties $\mu_2=\mu_1=1-\mu_3$, and our assumption on $\lambda_2$, it follows that the drift at the boundary is negative, driving the state to zero.

\begin{proof}
\par Consider the piecewise linear Lyapunov function defined by Eq.\ (17) and let $\mathcal{F}_t$ be the $\sigma$-field that corresponds to the history of the process until just before the arrivals at slot $t$; formally, $\mathcal{F}_t$  is the $\sigma$-algebra generated by $Q(0),A(0),\ldots,Q(t-1),A(t-1),Q(t)$. Throughout the proof we use $\mathbb{P}(X;A \mid \mathcal{H})$ and $\mathbb{E}[X; A \mid \mathcal{H}]$ to denote $\mathbb{P}(X \cdot 1_A \mid \mathcal{H})$ and $\mathbb{E}[X \cdot 1_A \mid \mathcal{H}]$, respectively, where $X$ is a random variable, $A$ is an event, and $\mathcal{H}$ is a $\sigma$-algebra on the sample space.

\par Our goal is to show that for sufficiently large (but fixed) $T \in \mathbb{N}$, there exist positive constants $\alpha$ and $\epsilon$, such that
\begin{equation}
\mathbb{E} \Big[ V(t+T) - V(t) + \epsilon\ ; \ V(t)>\alpha \ \Big| \ \mathcal{F}_t \Big] \leq 0,\nonumber
\end{equation}
and the desired result will follow from \cite{H82}.

\par Suppose that $V(t)> \alpha$. Then, we must have either $Q_2(t)>{\alpha}/{6} $ or $Q_3(t) > Q_1(t)+Q_2(t)+{\alpha}/{2}$. We will derive the desired drift inequality by considering separately these two cases. We assume that $T$ has been fixed to a suitably large value, and we define $\alpha$ by $\alpha=6 T$.

\bigskip{ \bf Case 1: $Q_2(t)>\alpha/6=T$.}
\\
Since at most one packet is removed at each time slot from queue 2, it is immediate that 
\begin{equation}
Q_2(\tau)>0, \qquad \forall \tau \in \{t,\ldots,t+T-1\}. \tag{18}
\end{equation}
Moreover, under the Max-Weight scheduling policy,
\begin{equation}
S_3(\tau) \cdot 1_{\{Q_3(\tau)>0\}} = S_3(\tau), \qquad \forall \tau \in \{t,\ldots,t+T-1\}. \tag{19}
\end{equation}
Eq. (19) implies that a service opportunity is never wasted in queue 3 throughout $\{t,\ldots,t+T-1\}$, assuming $Q_2(t)>T$. This is intuitively clear, because,  under the Max-Weight policy, queue 3 will not be served unless it becomes at least as large as queue 2,  hence positive, and a service opportunity will not be wasted).

Let
\begin{equation}
J(t)=Q_1(t)+Q_2(t)-Q_3(t). \nonumber
\end{equation}
Using this notation, and the queue length dynamics from Section 2, it can be verified that
\begin{align}
&\sum_{\tau=t}^{t+T-1} \Big( A_1(\tau)+A_2(\tau)-S_1(\tau) \cdot 1_{\{Q_1(\tau)>0\}}-S_2(\tau) \cdot 1_{\{Q_2(\tau)>0\}} \Big) \nonumber \\
= &\sum_{\tau=t}^{t+T-1} \Big( A_3(\tau)-S_3(\tau) \cdot 1_{\{Q_3(\tau)>0\}} \Big) + J(t+T)-J(t). \tag{20}
\end{align}
Moreover, the scheduling constraints  imply that
\begin{equation}
S_1(\tau)=S_2(\tau), \qquad \forall \tau \in \{t,\ldots,t+T-1\}. \tag{21}
\end{equation}
Furthermore, under the Max-Weight policy (or, in general, under any non-idling policy),
\begin{equation}
\sum_{\tau=t}^{t+T-1} (S_2(\tau)+S_3(\tau))=T. \tag{22}
\end{equation}
Eqs.\ (19) and (20) imply that
\begin{equation}
\sum_{\tau=t}^{t+T-1} (A_1(\tau)+A_2(\tau)-S_1(\tau)-S_2(\tau)) \leq \sum_{\tau=t}^{t+T-1} (A_3(\tau)-S_3(\tau)) + J(t+T)-J(t). \tag{23}
\end{equation}

\noindent Combining Eqs.\ (21) and (23), we have
\begin{equation}
\sum_{\tau=t}^{t+T-1} (A_1(\tau)+A_2(\tau)-2 \cdot S_2(\tau)) \leq \sum_{\tau=t}^{t+T-1} (A_3(\tau)-S_3(\tau)) + J(t+T)-J(t). \tag{24}
\end{equation}

\noindent Then, by taking into account Eq. (22), we get
\begin{equation}
-3 \cdot \sum_{\tau=t}^{t+T-1} S_2(\tau) \leq - \sum_{\tau=t}^{t+T-1} (1+A_1(\tau)+A_2(\tau)-A_3(\tau)) + J(t+T)-J(t). \tag{25}
\end{equation}

\par Let us now examine the implications of this inequality on the evolution of queue 2 throughout $\{t,\ldots,t+T-1\}$. We have
\begin{align}
3 (Q_2(t+T)-Q_2(t)) &= 3 \cdot \sum_{\tau=t}^{t+T-1} (A_2(\tau)-S_2(\tau) \cdot 1_{\{Q_2(\tau)>0\}}) \nonumber \\
&= 3 \cdot \sum_{\tau=t}^{t+T-1} (A_2(\tau)-S_2(\tau)) \nonumber \\
&\leq - \sum_{\tau=t}^{t+T-1} (1+A_1(\tau)-A_3(\tau)-2 A_2(\tau)) + J(t+T)-J(t), \nonumber
\end{align}

\noindent where the second equality follows from Eq. (18), and the inequality follows from Eq. (25). This implies that
\begin{align}
V(t+T)-V(t) = \ &3 (Q_2(t+T)-Q_2(t)) + [-J(t+T)]^+ - [-J(t)]^+ \nonumber \\
\leq &- \sum_{\tau=t}^{t+T-1} (1+A_1(\tau)-A_3(\tau)-2 A_2(\tau)) \nonumber \\
&+ J(t+T) - J(t) + [-J(t+T)]^+ - [-J(t)]^+ \nonumber \\
= &- \sum_{\tau=t}^{t+T-1} (1+A_1(\tau)-A_3(\tau)-2 A_2(\tau)) + [J(t+T)]^+ - [J(t)]^+. \nonumber
\end{align}
Therefore,
\begin{align}
\mathbb{E} \Big[V(t+T)-V(t)\ ; Q_2(t)>T \ \Big| \ \mathcal{F}_t \Big] \leq &-\delta \cdot T \cdot \mathbb{P} \Big(Q_2(t)>T\ \Big| \ \mathcal{F}_t \Big)  \nonumber \\
&+ \mathbb{E} \Big[ [J(t+T)]^+ - [J(t)]^+ ; Q_2(t)>T \ \Big| \ \mathcal{F}_t \Big]. \tag{26}
\end{align}
where $\delta=1+\lambda_1-\lambda_3-2 \lambda_2$ is positive by assumption.

\par We have also assumed that there exists $\gamma >0$, such that $\mathbb{E}[A_1^{1+\gamma}(0)]$ is finite. Based on this, we will prove that $\mathbb{E}\big[\, [J(t+T)]^+ - [J(t)]^+ \mid \mathcal{F}_t]\big]$ scales sublinearly in $T$. This, in turn, will imply that the right-hand side of Eq. (26) is negative, provided $T$ is sufficiently large. 

\par By disregarding the contribution of queue 3, we can bound from above $[J(\tau)]^+,\ \tau \in \{t,\ldots,t+T\},$ by the sum of the lengths of queues 1 and 2 during that interval. Moreover, whenever $[J(\tau)]^+$ is nonzero, both queues 1 and 2 are served at unit rate. Through Lindley's recursion and simple calculations, it can be verified that
\begin{equation}
[J(t+T)]^+ - [J(t)]^+ \leq \max_{1 \leq s \leq T} \Big\{ \sum_{\tau = t+T-s}^{t+T-1} (A_1(\tau)-1) \Big\}  + \max_{1 \leq s \leq T} \Big\{ \sum_{\tau = t+T-s}^{t+T-1} (A_2(\tau)-1) \Big\}. \tag{27}
\end{equation}

\noindent Also, the following inequality holds for all $i \in \{1,2\}$, and for all $s \in \{1,\ldots,T\}$:
\begin{equation}
\sum_{\tau = t+T-s}^{t+T-1} (A_i(\tau)-1) = (\lambda_i-1) s + \sum_{\tau = t+T-s}^{t+T-1} (A_i(\tau)-\lambda_i) \leq \Big| \sum_{\tau = t+T-s}^{t+T-1} (A_i(\tau)-\lambda_i) \Big|. \tag{28}
\end{equation}
Eqs. (27) and (28) imply that for any fixed $\gamma' \in (0,\gamma)$,
\begin{align}
\mathbb{P} \Big( [J(t+T)&]^+ - [J(t)]^+ \geq c\ ; Q_2(t)>T\ \Big| \ \mathcal{F}_t \Big) \nonumber \\
\leq &\mathbb{P} \Big( \max_{1 \leq s \leq T} \Big\{ \sum_{\tau = t+T-s}^{t+T-1} (A_1(\tau)-1) \Big\}  + \max_{1 \leq s \leq T} \Big\{ \sum_{\tau = t+T-s}^{t+T-1} (A_2(\tau)-1) \Big\} \geq c\ ; Q_2(t)>T\ \Big| \ \mathcal{F}_t \Big) \nonumber \\
\leq &\mathbb{P} \Big( \max_{1 \leq s \leq T} \Big\{ \Big| \sum_{\tau = t+T-s}^{t+T-1} (A_1(\tau)-\lambda_1) \Big| \Big\} \geq \frac{c}{2} \ ; Q_2(t)>T\ \Big| \ \mathcal{F}_t \Big) \nonumber \\
&+ \mathbb{P} \Big( \max_{1 \leq s \leq T} \Big\{ \Big| \sum_{\tau = t+T-s}^{t+T-1} (A_2(\tau)-\lambda_2) \Big| \Big\} \geq \frac{c}{2} \ ; Q_2(t)>T\ \Big| \ \mathcal{F}_t \Big) \nonumber \\
= &\mathbb{P} \Big( \max_{1 \leq s \leq T} \Big\{ \Big| \sum_{\tau = t+T-s}^{t+T-1} (A_1(\tau)-\lambda_1) \Big|^{1+\gamma'} \Big\} \geq \Big( \frac{c}{2} \Big)^{1+\gamma'} ; Q_2(t)>T\ \Big| \ \mathcal{F}_t \Big) \nonumber \\
&+ \mathbb{P} \Big( \max_{1 \leq s \leq T} \Big\{ \Big| \sum_{\tau = t+T-s}^{t+T-1} (A_2(\tau)-\lambda_2) \Big|^{1+\gamma'} \Big\} \geq \Big( \frac{c}{2} \Big)^{1+\gamma'} ; Q_2(t)>T\ \Big| \ \mathcal{F}_t \Big) \tag{29}
\end{align}

\par Notice that the sequence
\begin{equation}
\Big\{ \sum_{\tau = t+T-s}^{t+T-1} (A_i(\tau)-\lambda_i);\ s \in \mathbb{N} \Big\}, \qquad i \in \{1,2,3\}, \nonumber
\end{equation}
is a martingale. Consequently, the sequence
\begin{equation}
\Big\{\Big| \sum_{\tau = t+T-s}^{t+T-1} (A_i(\tau)-\lambda_i) \Big|^{1+\gamma'};\ s \in \mathbb{N} \Big\} \nonumber
\end{equation}
is a nonnegative submartingale. Doob's submartingale inequality (see, e.g., Section 14.6 of \cite{WI91}) and Eq.\ (29) imply that
\begin{align}
\mathbb{P} \Big( [J(t+T)]^+ - [J(t)]^+ &\geq c\ ; Q_2(t)>T\ \Big| \ \mathcal{F}_t \Big) \nonumber \\
\leq &\,\Big( \frac{2}{c} \Big)^{1+\gamma'} \cdot \mathbb{E} \Big[ \Big| \sum_{\tau = t}^{t+T-1} (A_1(\tau)-\lambda_1) \Big|^{1+\gamma'}\Big] \cdot 1_{Q_2(t)>T} \nonumber \\
+ &\Big( \frac{2}{c} \Big)^{1+\gamma'} \cdot \mathbb{E} \Big[ \Big| \sum_{\tau = t}^{t+T-1} (A_2(\tau)-\lambda_2) \Big|^{1+\gamma'}\Big] \cdot 1_{Q_2(t)>T}.  \tag{30}
\end{align}
Moreover, the Marcinkiewicz-Zygmund Strong Law of Large Numbers states that
\begin{equation}
\frac{\Big| \displaystyle{\sum_{\tau = t}^{t+T-1}} (A_i(\tau)-\lambda_i) \Big|}{T^{{1}/{(1+\gamma)}}} \stackrel{\mathcal{L}^1}{\longrightarrow} 0, \qquad i \in \{1,2,3\} \tag{31}
\end{equation}
(see, e.g., Chapter 6.10 of \cite{G05}). Eqs.\ (30) and (31) imply that  if $T$ is sufficiently large, then there exists $k>0$  (independent of $T$ and $c$), such that
\begin{equation}
\mathbb{P} \Big( [J(t+T)]^+ - [J(t)]^+ \geq c\ ;Q_2(t)>T\ \Big| \ \mathcal{F}_t \Big) \leq \frac{k}{c^{1+\gamma'}} \cdot T^{\frac{1+\gamma'}{1+\gamma}}
\cdot 1_{Q_2(t)>T},\nonumber
\end{equation}
for all $c \geq 0$. This gives
\begin{equation}
\mathbb{E} \Big[ [J(t+T)]^+ - [J(t)]^+ ; Q_2(t)>T\  \Big|\ \mathcal{F}_t \Big] \leq k \cdot T^{\frac{1+\gamma'}{1+\gamma}} \cdot \sum_{c=1}^\infty \frac{1}{c^{1+\gamma'}} \cdot 1_{Q_2(t)>T} . \nonumber
\end{equation}
Since $\gamma'>0$, the latter sum converges, so there exists $k'>0$ (independent of $T$), such that
\begin{equation}
\mathbb{E} \Big[ [J(t+T)]^+ - [J(t)]^+ ; Q_2(t)>T\  \Big| \ \mathcal{F}_t \Big] \leq k' \cdot T^{\frac{1+\gamma'}{1+\gamma}} \cdot 1_{Q_2(t)>T}. \tag{32}
\end{equation}
Finally, Eqs.\ (26) and (32), and the fact that $\gamma'<\gamma$, imply that for any fixed $\epsilon>0$, there exists a sufficiently large $T$, such that
\begin{equation}
\mathbb{E} \Big[V(t+T)-V(t)+\epsilon\ ; Q_2(t)>T\ \Big|\ \mathcal{F}_t \Big] \leq 0. \tag{33}
\end{equation}

\par \textbf{Remark:} the upper bound on $[J(\tau)]^+$ that we analyzed above can be viewed as an upper bound for a discrete-time stable M/GI/1 queue inwhere customers arrive in a Bernoulli fashion, and their service times are mutually independent and distributed identically to $A_1(0)+A_2(0)$. Since service times are heavy-tailed distributed, the expected steady-state workload in this queue is infinite (an immediate corollary of the Pollaczek-Khinchine formula). Combined with Fatou's lemma, this implies that the expected workload at time slot $t$ goes to infinity as $t$ increases. Eq.\ (32) shows that the expected workload goes to infinity at a sublinear rate (essentially, as $\mathcal{O}(T^{1/(1+\gamma)})$), assuming just the existence of the (1+$\gamma$) moment of service times. We note that Central-Limit-Theorem-type arguments cannot be used here, since the second moment of arrivals does not exist.

\bigskip{\bf Case 2: $Q_3(t)>Q_1(t)+Q_2(t)+3 T$.}\\ 
Since at most one packet is removed at each time slot from queue 3, it is immediate that 
\begin{equation}
Q_3(\tau)>0, \qquad \forall \tau \in \{t,\ldots,t+T-1\}. \nonumber
\end{equation}

\noindent Based on this, it can be easily verified that Eq. (25) still holds. This implies that
\begin{align}
3 (Q_2(t+T)-Q_2(t)) = \ &3 \cdot \sum_{\tau=t}^{t+T-1} (A_2(\tau)-S_2(\tau) \cdot 1_{\{Q_2(\tau)>0\}}) \nonumber \\
= \ &3 \cdot \sum_{\tau=t}^{t+T-1} (A_2(\tau)-S_2(\tau)) + 3 \cdot \sum_{\tau=t}^{t+T-1} S_2(\tau) \cdot 1_{\{Q_2(\tau)=0\}} \nonumber \\
\leq &- \sum_{\tau=t}^{t+T-1} (1+A_1(\tau)-A_3(\tau)-2 A_2(\tau)), \nonumber \\
&+ J(t+T) - J(t) + 3 \cdot \sum_{\tau=t}^{t+T-1} S_2(\tau) \cdot 1_{\{Q_2(\tau)=0\}}. \nonumber
\end{align}
Letting $D$ be the event $Q_3(t)>Q_1(t)+Q_2(t)+3 T$, this gives
\begin{align}
\mathbb{E} \Big[V(t+T)-V(t)\ ; D\ \Big| \ \mathcal{F}_t \Big] 
\leq &- \delta \cdot T \cdot 1_{D}  \nonumber 
+ \mathbb{E} \Big[ [J(t+T)]^+ - [J(t)]^+\ ; D\ \Big| \ \mathcal{F}_t \Big] \nonumber \\
&+ 3 \cdot \mathbb{E} \Big[ \sum_{\tau=t}^{t+T-1} S_2(\tau) \cdot 1_{\{Q_2(\tau)=0\}}\ ; D\ \Big| \ \mathcal{F}_t \Big]. \tag{34}
\end{align}
Working similar to the previous case, it can be verified that for any fixed $\gamma' \in (0,\gamma)$, there exists $k'>0$ (independent of $T$) such that
\begin{equation}
\mathbb{E} \Big[ [J(t+T)]^+ - [J(t)]^+\ ; D\ \Big| \ \mathcal{F}_t \Big] \nonumber \\
\leq k' \cdot T^{\frac{1+\gamma'}{1+\gamma}} \cdot 1_D. \tag{35}
\end{equation}

In view of Eqs.\ (34) and (35) and in order to establish the negative drift property of the Lyapunov function, it is sufficient to consider a large time horizon $T$, and to show that $\mathbb{E}[ \sum_{\tau=t}^{t+T-1} S_2(\tau) \cdot 1_{\{Q_2(\tau)=0\}} \mid \mathcal{F}_t]$ scales sublinearly in $T$.

\par In order to have a wasted service opportunity at queue 2, the schedule $\{1,2\}$ must first claim the server. This can only happen if $\sum_{\tau=t}^{t+T-1} (A_1(\tau)+A_2(\tau))$, the aggregate arrivals to queues 1 and 2 during the interval $\{t,\ldots,t+T-1\}$ exceed the initial difference between the weights of the two schedules (which is at least $3T$), minus the departures from queue 3 during the same period (which are at most $T$). It follows that
\begin{align}
\Big\{ \sum_{\tau=t}^{t+T-1} S_2(\tau) \cdot 1_{\{Q_2(\tau)=0\}} >0 \Big\} 
\subset &\Big\{ \sum_{\tau=t}^{t+T-1} (A_1(\tau)+A_2(\tau)) > 3 T -T \Big\} \nonumber \\
= &\Big\{ \sum_{\tau=t}^{t+T-1} (A_1(\tau)+A_2(\tau)-\lambda_1-\lambda_2) > (2-\lambda_1-\lambda_2) T \Big\} \nonumber \\
\subset &\Big\{ \Big| \sum_{\tau=t}^{t+T-1} (A_1(\tau)+A_2(\tau)-\lambda_1-\lambda_2) \Big| > (2-\lambda_1-\lambda_2) T \Big\}. \nonumber
\end{align}
Note that $\lambda_1+\lambda_2<2$, since the arrival rate vector was assumed to be in the stability region of the system. Then, using Markov's inequality and Eq.\ (34), we obtain that there exists $\xi>0$ (independent of $T$) such that
\begin{align}
&\mathbb{P} \Big( \sum_{\tau=t}^{t+T-1} S_2(\tau) \cdot 1_{\{Q_2(\tau)=0\}} >0\ ; D\ \Big| \ \mathcal{F}_t \Big) \nonumber \\
\leq &\mathbb{P} \Big( \Big| \sum_{\tau=t}^{t+T-1} (A_1(\tau)+A_2(\tau)-\lambda_1-\lambda_2) \Big| > (2-\lambda_1-\lambda_2) T\ ; D \ \Big| \ \mathcal{F}_t \Big) \nonumber \\
\leq &\frac{\mathbb{E} \Big[ \Big| \sum_{\tau=t}^{t+T-1} (A_1(\tau)+A_2(\tau)-\lambda_1-\lambda_2) \Big| \Big]}{(2-\lambda_1-\lambda_2) T}\, \cdot \, 1_D \nonumber \\
\leq &\xi \cdot T^{\frac{1}{1+\gamma}-1}\, \cdot \, 1_D.  \nonumber
\end{align}
provided $T$ is sufficiently large. 
Since the number of wasted service opportunities at queue 2 during an interval of length $T$ is bounded by $T$, we conclude that
\begin{equation}
\mathbb{E} \Big[ \sum_{\tau=t}^{t+T-1} S_2(\tau) \cdot 1_{\{Q_2(\tau)=0\}} \ ; D\ \Big| \ \mathcal{F}_t \Big] 
\leq T\cdot \xi \cdot T^{\frac{1}{1+\gamma}-1} \cdot 1_D=
\xi \cdot T^{\frac{1}{1+\gamma}} \cdot 1_D
.\nonumber\tag{36}
\end{equation}
Eqs. (34)-(36) imply that for any given $\epsilon>0$, there exists a sufficiently large $T$, such that
\begin{equation}
\mathbb{E} \Big[V(t+T)-V(t) + \epsilon\ ; D\ \Big| \ \mathcal{F}_t \Big] \leq 0. \tag{37}
\end{equation}
This completes the derivation of the drift inequality for the second case.

To summarize, we have shown, in Eqs.\ (33) and (37), that for any given $\epsilon>0$, we can choose $T$ sufficiently large and let $\alpha =6T$ to guarantee that
\begin{equation}
\mathbb{E} \Big[V(t+T)-V(t)+\epsilon\ ; V(t)>\alpha\ \Big| \ \mathcal{F}_t \Big] \leq 0, \qquad t \in \mathbb{Z}_+. \nonumber
\end{equation}
Theorem 2.3 of \cite{H82}, and the fact that the sequence $\{V(t);\ t \in \mathbb{Z}_+\}$ converges in distribution, imply that the steady-state length of queue 2, $Q_2$, is exponential-type, and, in particular, has finite expectation.
\end{proof}

\medskip\par An immediate corollary of Proposition 3 and Lemma 2 is that queue 2 is delay stable, provided its arrival rate is sufficiently small. To establish this, we assumed that the stochastic processes $\{A_2(t);\ t \in \mathbb{Z}_+\}$ and $\{A_3(t);\ t \in \mathbb{Z}_+\}$ are exponential-type. At the same time, this additional assumption enabled us to prove a result that is much stronger than delay stability: combining Proposition 3 with the distributional Little's Law (e.g., see \cite{W91}), we have that \textbf{the steady-state delay in queue 2 is exponential-type}.


\bigskip\section{Discussion}

\par We considered a simple queueing system with heavy-tailed traffic, and showed a rate-dependent delay stability phenomenon under the Max-Weight scheduling policy: there is a part of the stability region where a certain queue is delay stable, and a part of the stability region where the same queue is delay unstable. We note that this phenomenon would not arise if all stochastic primitives were light-tailed.

\par Despite the simplicity of the underlying queueing system, the proofs of our main results (Propositions 2 and 3) are long, somewhat technical, and, in some sense, ``tailored'' to the specifics of the system. Hence, generalizing these results using similar methods is not straightforward. What is missing is a systematic and practical methodology for determining which queues are delay stable, which are delay unstable, and over which parts of the stability region, for any given switched queueing system. 

\par On a more technical side, the proof of the delay stability result (Proposition 3) requires the additional assumption that all light-tailed arrival processes are exponential-type. Only then is drift analysis of a piecewise linear Lyapunov functions meaningful (i.e., we can invoke the results in \cite{H82}). One possible way to overcome this restriction would be through a suitably defined piecewise quadratic Lyapunov function. However, drift analysis seems cumbersome in that case.


\end{document}